\documentclass{ifacconf}

\usepackage{graphicx}      
\usepackage{natbib}        

\usepackage{enumitem}
\usepackage{amssymb}
\usepackage{amstext}
\usepackage{amsmath}
\usepackage[ruled,vlined,algo2e]{algorithm2e}
\usepackage{booktabs}
\usepackage{ductri}
\usepackage{cancel}

\usepackage{float}

\begin{document}
\begin{frontmatter}

\title{PSO-Based Adaptive NMPC for Uranium Extraction-Scrubbing Operation in Spent Nuclear Fuel Treatment Process \thanksref{footnoteinfo}} 

\thanks[footnoteinfo]{The authors thank ORANO for partial financial support for the project.}

\author[Second]{Duc-Tri Vo$^{*,}$} 
\author[First]{Ionela Prodan} 
\author[First]{Laurent Lefèvre}
\author[Second]{Vincent Vanel} 
\author[Second]{Sylvain Costenoble}
\author[Second]{Binh Dinh}

\address[First]{Univ. Grenoble Alpes, Grenoble INP, LCIS, 26000, Valence, France (e-mail: \{duc-tri.vo, ionela.prodan, laurent.lefevre\}@lcis.grenoble-inp.fr).}
\address[Second]{CEA, DES, ISEC, DMRC, Univ Montpellier, Marcoule, France (e-mail: \{vincent.vanel, sylvain.costenoble, binh.dinh\}@cea.fr)}

\begin{abstract}                
This paper addresses the particularities of adaptive optimal control of the uranium extraction-scrubbing operation in the PUREX process. The process dynamics are nonlinear, high dimensional, and have limited online measurements. In addition, analysis and developments are based on a qualified simulation program called PAREX, which was validated with laboratory and industrial data. The control objective is to stabilize the process at a desired solvent saturation level, guaranteeing constraints and handling disturbances. The developed control strategy relies on optimization-based methods for computing control inputs and estimates, i.e., Nonlinear Model Predictive Control (NMPC) and Nonlinear Moving Horizon Estimation (NMHE). The designs of these two associated algorithms are tailored for this process's particular dynamics and are implemented through an enhanced Particle Swarm Optimization (PSO) to guarantee constraint satisfaction. Software-in-the-loop simulations using PAREX show that the designed control scheme effectively satisfies control objectives and guarantees constraints during operation.
\end{abstract}


\end{frontmatter}

\section{Introduction}\label{sec:introduction}
\subsection{Motivation}
Spent nuclear fuels from nuclear reactors predominantly comprise 95\% uranium and 1\% plutonium, both of which are amenable to recovery. The remaining 4\%, characterized by its pronounced radioactivity, is classified as the ultimate waste. The PUREX process, denoting ``Plutonium, Uranium, Reduction, EXtraction," was conceived to effectuate the recovery and purification of uranium and plutonium from spent nuclear fuels (\cite{Vaudano2008}). It finds application at the La Hague nuclear fuel reprocessing plant in northern France, owing to its ability to recover these materials with an elevated degree of purity. This procedural approach facilitates the recycling of uranium and plutonium to produce fresh fuel resources while curtailing the volume of ultimate waste. Therefore, the PUREX process advances sustainable fuel cycles and nuclear waste management paradigms.

One of the primary objectives of the PUREX process is to attain a notable recovery efficiency for uranium and plutonium while concurrently achieving a high decontamination factor of those elements toward the fission products. This objective, performed to several liquid-liquid extraction steps, can be realized through the deliberate manipulation of solvent saturation, accomplished by modulating the flow rate of the feed solution, i.e., the quantity of uranium introduced into the process can be increased or decreased as necessary. This study directs its attention to the extraction-scrubbing operation, which is the first step within the extraction cycles and substantially influences the overall efficiency of the process.

\subsection{PUREX Modelling: The PAREX Code}

The French Alternative Energies and Atomic Energy Commission (CEA) developed a simulation software, PAREX, which has the ability to emulate the extraction operations within the PUREX process. This computational tool was validated through rigorous comparison with laboratory and industrial-scale experimental data, affirming its proficiency in effectively replicating PUREX process operations under transient and steady-state conditions, as reported by \cite{Bisson2016}. PAREX finds application at the La Hague facility, aiming to optimize processes, troubleshoot operational issues, and conduct safety analyses (\cite{Bisson2016}). Within this study, the PAREX code is the primary tool employed for system dynamics analysis and the formulation of control strategies. Since the process dynamics are complicated, we aim to integrate the PAREX simulator directly into the control scheme (as the predictor and the estimator). This way, we can benefit from the qualified mathematical model and the efficient, robust numerical solver employed in PAREX.

\subsection{Literature Review}
To the best of our knowledge, the research landscape concerning this specific process control is notably limited. A study within the same domain is reported in \cite{vo2023}, wherein a tracking Nonlinear Model Predictive Control (NMPC) strategy was designed to maximize process production. The simulation outcomes, both under nominal and perturbed conditions, demonstrated the efficacy of NMPC in stabilizing the process at its optimal working condition. However, this work used a simplified model for both NMPC developments and process simulation, which had errors compared to the qualified model in PAREX and the actual process. In addition, all the states and disturbances were assumed to be perfectly measured, which is a strong assumption compared to practical implementation conditions. Finally, to broaden the view, studies on the control of similar processes can be found in \cite{vo2023} and the references therein.

Our particular control problem is nonlinear and subject to hard constraints. Thus, Model Predictive Control, a well-known control method in academia and industry (\cite{Qin2003}), is a promising candidate. The basic idea of MPC is to find the control inputs in a prediction horizon that minimize a cost function. In other words, MPC solves an optimization problem online to obtain optimal control inputs. For tracking problems, the cost function is usually chosen to represent the tracking error in the prediction horizon. However, another challenge in this study stems from the limited availability of online measurements: only one state is measured online. The choice of measurements is made by considering sensitivity analysis (to have measurements on the most sensitive points), type of device (pulsed column, centrifugal extractor, mixer-settler), and cost. It is possible to carry out several measurement points online, but this study focuses on the case with one measurement.

Consequently, there arises a need for an estimator to compute estimations of states and parameters using the available online measurements. These estimated values are subsequently integrated into the NMPC as feedback, constituting a configuration commonly denoted as ``output feedback MPC" or ``adaptive MPC" (\cite{Jabarivelisdeh2020, Valipour2022}). A promising candidate for this purpose is the optimization-driven Moving Horizon Estimation (MHE) method (\cite{Rawlings2013}). MHE estimates states and parameters simultaneously by minimizing the errors between nominal outputs and online measurements. It has widespread application across diverse domains, particularly within process control (\cite{LiuJinfeng2018,Tuveri2023}). MPC and MHE theory, computation, and implementation are discussed systematically and comprehensively in \cite{James2022}.

Implementing MPC and MHE involves solving their optimization problems. Thus, an optimization algorithm is required. Particle Swarm Optimization (PSO) is a derivative-free, population-based optimization algorithm inspired by natural collective behavior observed in birds and fish, as initially proposed by \cite{Kennedy1995}. Within the PSO framework, each particle position is represented as a vector containing a candidate solution for the optimization problem. Initially, particles are distributed in the search space. Then, a cost value is attributed to each particle based on their position. Subsequently, particles undergo iterative movements in pursuit of the optimal position associated with the lowest cost. Successful applications of PSO in nonlinear control and estimation are reported in \cite{Schwaab2008, Zietkiewicz2020, Gbadega2022}. An essential feature of PSO is its independence from the necessity of function gradients, thereby obviating the need for an explicit analytical representation of the cost function and system dynamics. This characteristic aligns with our control objectives, where all computations are based on the ``black box" PAREX. 

\subsection{Contributions and Paper Organization}
This paper introduces a non-linear adaptive control strategy tailored to the PUREX process's uranium extraction-scrubbing operation. The proposed control approach combines non-linear Model Predictive Control, Moving Horizon Estimation, and Particle Swarm Optimization. Notably, the control strategy relies on the qualified simulation software PAREX as the primary computational instrument, as no explicit mathematical model equations is required. Briefly, our main contributions are:
\begin{itemize}
    \item proposing an adaptive control strategy based on NMPC and MHE to account for constraints satisfaction (e.g., uranium concentration in the fission product, equipment limits), setpoint tracking (i.e., aqueous uranium concentration at stage 9 settler), and parameters uncertainty (i.e., variation in the fresh solvent flow rate);
    \item proposing an enhanced PSO implementation for both the NMPC and MHE algorithms that can efficiently handle constraints on states, overshoots, control inputs, rate of changes of control inputs, and parameter estimation;
    \item study the effectiveness of the proposed control strategy through software-in-the-loop (SIL) testing using the validated PAREX simulator developed by CEA.
\end{itemize}

The paper is organized as follows. The process dynamics and the control problem setup is presented in Section 2. Then, the control strategy for the control problem is developed in Section 3. Section 4 presents our enhanced PSO algorithm to solve the optimization problems in the control strategy. Simulation results of nominal and disturbed case are presented in Section 5. Finally, conclusions are stated in Section 6.

\textbf{Notations:} Vectors are denoted by bold lowercase letters, and matrices are denoted by capital letters. $\bI,\, \mathbf{0}$ denotes identity and zero matrices of appropriate dimensions. $\bx_n$ denotes the $n^\text{th}$ element of $\bx$. $\bx_{n:m}$ denotes a vector slice from the $n^\text{th}$ to $m^\text{th}$ (included) elements of $\bx$. $\left\|\bx\right\|_\bQ := \bx^T\bQ\bx$. $\mathbb{N}_{m:n}:=\set{i \in \mathbb{N}| m\le i < n}$. $|\mathcal{C}|$ denotes the cardinality of the set $\mathcal{C}$. $U(\ba, \bp)$ denotes a uniformly random distribution within bounds $\ba,\, \bp$. Inequalities between vectors are element-wise.

\section{Uranium Extraction-Scrubbing Operation}
\begin{figure}[H]
    \centering
    \includegraphics[width=\linewidth]{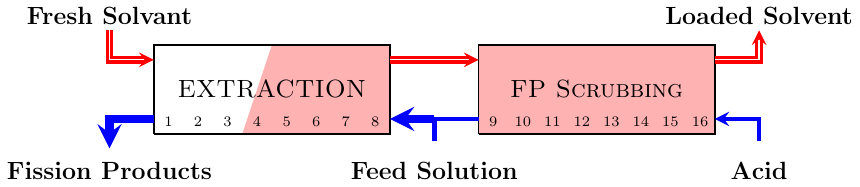}
    \caption{Extraction - fission products (FP) scrubbing step (\cite{vo2023}).}
    \label{fig:system_simplified_EN}
\end{figure}

Fig.~\ref{fig:system_simplified_EN} illustrates the fundamental architecture of our process, characterized by a sequential arrangement of 16 mixer-settler stages. The mathematical model within the PAREX platform was developed following a similar approach as the simplified one in the work by \cite{vo2023}. The main difference is the inclusion of specific experimentally derived parameters. Consequently, we adopt the discretized state space representation \eqref{eq:state_space} to derive the control strategy. It is worth noting that while the definition of $\ff$ is not explicitly given, its evaluation can be obtained utilizing PAREX.

\begin{subequations}
    \begin{gather}
    \bx(k+1) = \ff(\bx(k),\, u(k),\, q(k)),\\
    y(k+1) = \bx_{25},
\end{gather}\label{eq:state_space}
\end{subequations}
where:
\begin{itemize}
    \item $\bx \in \mathbb{R}_{128\times 1}$, uranium and acid concentrations in aqueous and organic phases of each stage, are system states;
    \item $u = \An{F}$, the feed solution flow rate, is the manipulated variable;
    \item $q = \On{E}$, the solvent flow rate, is the unknown disturbance. This type of disturbance can appear when there is a failure of mechanical coupling, pumps or flow meters in the system;
    \item $y = \UaqD{9}$, the aqueous uranium concentration at stage 9, is the controlled variable. This choice is based on the fact that this variable can effectively indicate the solvent saturation level, and its value is compatible with the uranium concentration sensor's operation range.
\end{itemize}




\subsubsection{Multiple time scales:} The control problem has to handle tasks at different time scales: i) first, it is necessary to read the measurements, which should be done \textit{online} with sufficient small sampling time, $T_\text{meas}$, to obtain relevant feedback from the system;  ii) second, is to detect any system disturbances by comparing measurements with simulation results. This task is done with an estimation sampling time $T_\text{esti}$; iii) lastly, it is necessary to find the system's optimal control inputs, which generally requires the highest computation time. In addition, since the process dynamics is slow, the control sampling time $T_\text{ctrl}$ can be chosen as the largest sampling time, i.e., $T_\text{ctrl} \ge T_\text{esti} \gg T_\text{meas}$. For convinience in the discrete control scheme, we denote $N_\text{ctrl}=T_\text{ctrl}/T_\text{meas}$ and $N_\text{esti}=T_\text{esti}/T_\text{meas}$.

\subsection{Solvent Saturation and Desired Set-point}
\begin{figure}[ht]
    \centering
    \includegraphics[width=0.9\linewidth]{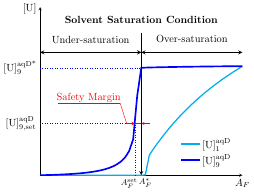}
    \caption{Steady state relationship of feed solution flow rates and uranium concentrations.}
    \label{fig:SSC}
\end{figure}

Saturation is a non-linear characteristic of this process dynamics, as illustrated in Fig.~\ref{fig:SSC}. Once the system is over-saturated, i.e., $A_F > A_F^*$, $\UaqD{1}$ increases drastically. Thus, uranium will go into extraction raffinates, which must be avoided. In addition, the system can quickly become over-saturated with a smaller safety margin, especially with uncertainties. Therefore, keeping the system working in the under-saturation region but close to the critical condition is desirable. The choice of the set point depends on the expected performances (recovery rate, decontamination factor, the specifics of the chemical system (extractant, media, concentration), and the feed solution composition). Latter in the simulations, we choose $\UaqD{9,\text{set}}=0.5 [U]^\text{aqD*}_9$ as illustrated in Fig.~\ref{fig:SSC}.

\subsection{Problem Statement}\label{sec:problem_statement}
The control problem is stated as follows: based on the PAREX Code and the state space representation \eqref{eq:state_space}, develop a control system that stabilizes $y=\UaqD{9}$ at a desired set point $y_\text{set}=\UaqD{9,\text{set}}$ in the presence of disturbance. Additionally, constraint on uranium concentration in the extraction raffinates $\UaqD{1}$ \eqref{eq:cons_uaqD1}, overshoots OS \eqref{eq:cons_OS}, control inputs $A_F$ \eqref{eq:cons_u}, and rates of control inputs \eqref{eq:cons_du} must be guaranteed during operation. Note that $\Delta_{A_F}^\text{max}$ is the maximum allowable variation in $A_F$ and \eqref{eq:cons_du} is considered as a soft constraint and can be relaxed to guarantee the hard constraints \eqref{eq:cons_uaqD1}-\eqref{eq:cons_u}:
\begin{subequations}
    \begin{gather}
        \UaqD{1} \le \UaqD{1,\text{tol}},\label{eq:cons_uaqD1}\\
        \text{OS} = y/y_\text{set} - 1 \le \text{OS}^\text{max}, \label{eq:cons_OS}\\
        A_F^\text{min} \le A_F \le A_F^\text{max}\label{eq:cons_u},\\
        \left| A_F(k+1) - A_F(k) \right|\le \Delta_{A_F}^\text{max}. \label{eq:cons_du}
    \end{gather}
\end{subequations}

Furthermore, in our control problem, the steady tracking error $e$ is not necessarily zero, which is hard to achieve in the presence of uncertainties. We accept that the system is stabilized if the tracking error is below a predefined tolerance \eqref{eq:steady_tolerance}, where $\varepsilon_\text{ss}\in (0,1)$ is a relative factor:
\begin{gather}
    e = \left| y - y_\text{set}\right| \le \varepsilon_\text{ss} y_\text{set}.\label{eq:steady_tolerance}
\end{gather}
As we will see, this choice helps reduce MPC's computational cost and, thus, makes the practical implementation of the control strategy more efficient.

\section{Control Strategy}
\begin{figure}[H]
    \centering
    \includegraphics[width=\linewidth]{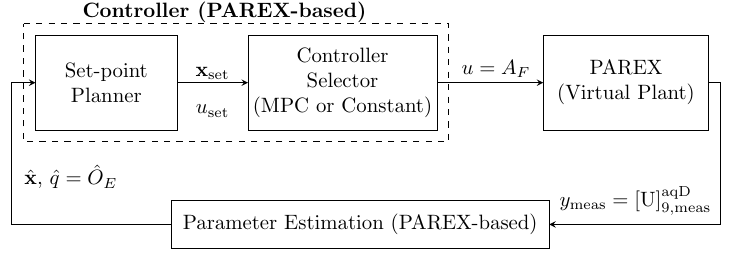}
    \caption{Proposed MPC-MHE control strategy using the PAREX simulator in the loop.}
    \label{fig:control_strategy}
\end{figure}
This section presents the developed control strategy illustrated in Fig.~\ref{fig:control_strategy}. First, the online measurement $y_\text{meas}$ is read and then sent to the Moving Horizon Estimator (MHE) for state and parameter estimations. After that, estimated states and parameters are used to compute the control input $u$. We have a controller selector inside the controller block that allows us to turn off MPC and use constant control inputs under some conditions. This behavior allows us to reduce computation costs while adhering to control objectives and constraints.


\subsection{MHE for Parameter Estimation}

As shown in Fig.~\ref{fig:control_strategy} we delineate in the following the MHE procedure for estimating the solvent flow rate, that is, the $O_E$ parameter in \eqref{eq:state_space} and then compute the estimated state values. This approach avoids solving a large-scale optimization problem since the system is already high dimensional with 128 states. Firstly, denote $e_\text{esti}$ as the estimation errors, i.e., the deviation between the measurements $y_\text{meas}$, and the estimated value $\hat{y}$:
\begin{gather}
    e_\text{esti} = y_\text{meas} - \hat{y}. \label{eq:esti_error}
\end{gather}
Note that, $y = \UaqD{9}$ is the aqueous uranium concentration at stage 9. Denote $\mathcal{K}^E= \set{k_{-N_e},\, \dots,\, k}$ as the set of co-incident points at which we want to minimize $e_\text{esti}$. Regarding the set $\mathcal{K}^E$, at the first time step $k_{-N_e}$, the state value must be good enough since $\bx\left(k_{-N_e}\right)$ is the initial condition for the optimization problem \eqref{eq:optimization_mhe}. In this work, we choose $k_{-N_e}$ corresponding to the latest window of size $N_w$: $\left[k_{-N_e}-N_w,\, k_{-N_e}\right]$ in which $e_\text{esti}$ is always below a tolerance $\varepsilon_{-N_e}$:
    \begin{gather}
    \max_{i\in [k_{-N_e}-N_w, k_{-N_e}]} \left| e_\text{esti} \right| \le \varepsilon_{-N_e}. \label{eq:condition_k0}
    \end{gather}
Denote $k_{-1}:= k $ as the rest of the time steps, i.e. $k_{-N_e+1},\,\dots,\,k_{-2}$, which are evenly spaced samples on the interval $\left[k_{-N_e},\, k_{-1}\right]$. Hence, the estimated parameter $\hat{q}(k)$ can be obtained by solving the following optimization problem, $\forall i \in \mathcal{K}^E_k$:
\begin{subequations}
\begin{gather}
    \min_{\{\hat{q}(i)\}_{i \in \mathcal{K}^E_k}} \sum_{i \in \mathcal{K}^E_k} \alpha_i e_\text{esti}^2,
\end{gather}
subject to:
\begin{gather}
        \bhx(i+1) = \ff(\bhx(i), u(i), \hat{q}(i)),\\
        \hat{y}(i) = \bhx_{25}(i),
\end{gather}\label{eq:optimization_mhe}
\end{subequations}
where $\alpha_i$ is the weighting factor and $\mathcal{K}^E_k$ is the set of co-incident points at which $e_\text{esti}$ is minimized. Putting high weighting factors on the latest co-incident points is preferable since the latest feedback is much more important.

One observation is that estimating $q=O_E$ by solving \eqref{eq:optimization_mhe} is costly, and the computational cost can be reduced by setting $\hat{q}(k) = \hat{q}(k-1)$ if the following condition holds:
\begin{gather}
    \left|e_\text{esti}\right| \le \varepsilon_\text{active}, \label{eq:mhe_active}
\end{gather}
where $\varepsilon_\text{active}$ represents the acceptable deviation between the estimated value and online feedback. 

The algorithm for implementing the parameter estimation is briefly summarized as follows. First, we read the measurements up to time $k$. After, we check if \eqref{eq:mhe_active} holds so that we can set $\hat{q}(k) = \hat{q}(k-1)$. Otherwise, we compute $\hat{q}$ using the MHE method, which includes constructing $\kappa^E$ and solving \eqref{eq:optimization_mhe}. The estimated state corresponding to $\hat{q}$ is then obtained using PAREX.


\subsection{Controller Design} 
This subsection introduces the set-point tracking NMPC and the controller selector as shown in Fig \ref{fig:control_strategy}. Firstly, PAREX is used to find the set point $\left(\bx_\text{set}, u_\text{set}\right)$ by solving the steady-state equation, where $y_\text{set}=\bx_\text{set,25}$:
\begin{gather}
    \left.\ff(\bx(k), u(k), \hat{q}(k))\right|_{y(k) = y_\text{set}} = 0 \label{eq:steady_state}.
\end{gather}

The control sampling time $T_\text{ctrl}$ is significantly larger than the measurement sampling time $T_\text{meas}$. Hence, to describe the MPC algorithm, denote $\mathcal{K}^C$ the set of time steps at which we compute the control input using MPC:
\begin{gather}
    \mathcal{K}^C := \set{k | \left(k \mod N_\text{ctrl}\right) = 1, \, \forall k\in \mathbb{N}_{1:N_\text{sim}}} \label{eq:K_MPC},
\end{gather}
where $N_\text{sim}$ denotes the total simulation time. Then, the set of control time steps in the prediction horizon at time step $k$, $\mathcal{K}^C_{k}$, is given as 
\begin{gather}
    \mathcal{K}^C_{k} := \set{k_p \in \mathcal{K}^C \;|\; k \le k_p \le k+N_pN_\text{ctrl}}, \label{eq:K_MPC_pred}
\end{gather}
with $N_p$ denoting the prediction horizon.
At the control time step $k \in \mathcal{K}^C$, given the estimated initial state $\bhx(k)$, denote $\bx(i|k)$ the predicted state at time step $j$ driven by the predicted control inputs $\{u(j|k)\}_{j=k}^{i}$, $\forall i \in \mathbb{Z}^+$, $i>k$. Assuming that the control input $u$ is constant over the interval $\left[jk, jk+N_\text{ctrl} -1\right]\; \forall j \in \mathbb{N}_{0:N_p-1}$ and denoting the tracking error for the state by $\btx := \bx - \bx_\text{set}$ and for the input by $\tilde{u}:=u - u_\text{set}$, then we define the quadratic cost function as:
\begin{align}
&\ell_\text{MPC}=\sum_{i\in\mathcal{K}^C_k}\left(\left\|\btx(i|k)\right\|_{\bQ}^2 + \left\|\tilde{u}(i|k)\right\|_{\bR}^2 \right) \notag\\
    &+  \sum_{i\in\mathcal{K}^C_k}\left\|u(i|k) - u(i-1|k)\right\|_{\bS}^2 + \left\|\btx(N+1|k)\right\|_{\bP}^2, \label{eq:cost_function}
\end{align}
where $\bQ$, $\bR$, $\bP$, $\bS$ denote symmetric positive definite weighting matrices of appropriate dimensions. The optimization problem is defined as:
\begin{subequations}
    \begin{align}
\min_{\{u(i|k)\}_{i\in \mathcal{K}^C_k}} \ell_\text{MPC},
\end{align}
subject to: 
\begin{gather}
    \bx(j+1|k) = \ff(\bx(j|k),u(j|k),\hat{q}(k)),\\
    \bx(k|k) = \bhx (k),\\
    u(k-1|k):=u(k-1),\\
    \bx(i|k) \ge 0, \label{eq:mpc_x}\\
    \UaqD{1}(i|k) = \bx_{17} (i|k) \le \UaqD{1,\text{tol}}, \label{eq:mpc_uaqD1}
\end{gather}
\begin{gather}
    \text{OS}(i|k) \le y(i|k)/y_\text{set} - 1 \le \text{OS}^\text{max}, \label{eq:mpc_os}\\
    A_F^\text{min} \le u(i|k) \le A_F^\text{max}, \label{eq:mpc_u}\\
    \left|u(i|k) - u(i-1|k) \right|\le \Delta A_F^\text{max}.\label{eq:mpc_du}
\end{gather}\label{eq:optimization_mpc}
\end{subequations}


To reduce the computation cost of NMPC, we propose the \textit{Controller Selector}, which allows to switch to a constant controller $u(k)=u_\text{set}$ without solving the NMPC optimization problem. The switching to the constant controller is done if by predicting over the next $N_{p2}$ time steps, the condition \eqref{eq:steady_tolerance} holds, i.e.:
\begin{subequations}
    \begin{gather}
    \left|y(i|k) - y_\text{set}\right| \le \varepsilon_\text{ss} y_\text{set}, \forall i \in \mathcal{K}^C_{k2}, \\
    \mathcal{K}^C_{k2} := \set{k_p \in \mathcal{K}^C \;|\; k \le k_p \le k+N_{p2}N_\text{ctrl}}. \label{eq:K_MPC_pred2}
\end{gather}\label{eq:controller_selector}
\end{subequations}
From our experience, $N_{p2}$ is preferred to be much larger than $N_p$ to ensure that the system is about to be steady.

The controller design is summarized as follows: We first get the estimated values from the MHE. If condition \eqref{eq:controller_selector} holds, we can consider  $u(k)=u_\text{set}$. Otherwise we obtain $u(k)$ by solving the NMPC optimization problem \eqref{eq:optimization_mpc}.

  


\section{Particle Swarm Optimization\\for the proposed MPC and MHE designs}\label{sec:appendix}    

With the PAREX simulator integrated into the control loop as in Fig. \ref{fig:control_strategy}, we propose the use of PSO to solve the NMPC and MHE optimization problems developed in the previous section. The advantage is that PSO does not require an explicit declaration of the mathematical equations of the model; it allows the use of the ``black box" PAREX to solve optimization problems. In this section, we developed and extended the PSO formulation to accommodate the NMPC and MHE designs. The PSO implementation will be analyzed here for the MPC optimization problem \eqref{eq:optimization_mpc} and can be similarly applied for solving the MHE optimization problem \eqref{eq:optimization_mhe}. Since multiple notations are used, we will first describe the framework of the PSO algorithm in Sec.~\ref{sec:basic_pso}, then present our extensions later in Sec.~\ref{sec:extended_pso}.

\subsection{Global-best Guaranteed Convergence PSO}\label{sec:basic_pso}
\subsubsection{Basic Global-best PSO}
Assume that the swarm has $N_\text{part}$ particles. At iteration $i$, denote the $n^{\text{th}}$ particle's position and cost by $\bp_n^{(i)}$, and $\ell_n^{(i)} := \ell(\bp_n^{(i)})$ respectively. In addition, denote $\bbp^{(i)}$, $\bbell_n^{(i)}$ the \textit{personal best} position and cost of the $n^\text{th}$ particle; $\bbbp^{(i)}$, $\bbbell_n^{(i)}$ the \textit{global best} position and cost of the swarm:
\begin{subequations}
    \begin{gather}
        \bbp_{n}^{(i)} := \argmin_{\substack{\bp = \bp_{n}^{(j)},\, j\in \mathbb{N}_{0:i+1}}} \ell\left(\bp\right), \quad \bbell_n^{(i)} := \ell\left(\bbp_{n}^{(i)}\right),\\
        \bbbp^{(i)} := \argmin_{\substack{\bp = \bp_{n}^{(j)},\\j\in \mathbb{N}_{0:i+1},\,n\in \mathbb{N}_{0:N\text{part}}}} \ell\left(\bp\right), \quad \bbbell_n^{(i)} := \ell\left(\bbbp_{n}^{(i)}\right).
    \end{gather}\label{eq:pb_gb_position}
\end{subequations}
At each iteration, particle positions are updated with velocity velocities $\bv_n^{(i)}$ given in \eqref{eq:pso_p_update}. Particles velocities are computed using \eqref{eq:pso_v}, with $\br_1^{(i)},\, \br_2^{(i)} \sim U(\mathbf{0},\mathbf{1})$, $w^{(i)} \in (0,1)$ is the inertia weight factor, and $c_1^{(i)},\, c_2^{(i)}>0$ are acceleration constants. In this work, the linear decreasing inertia weight and time-varying acceleration constants \eqref{eq:pso_w}-\eqref{eq:pso_c2} are used, with $w_i=0.9$, $w_f=0.4$, $c_{1i}=c_{2f}=2.5$, $c_{1f}=c_{2i}=0.5$ as suggested in \cite{Naka2001} and \cite{Ratnaweera2004}. Note that the velocity should be clamped within a predefined range to avoid large updates that make the particle go out of the search space or miss a potential search region. Initially, particle positions are scattered \textit{uniformly} in the search space, and velocities are set to zero.
\begin{subequations}
\begin{align}
    \bp_n^{(i+1)} &= \bp_n^{(i)} + \bv_n^{(i)}, \label{eq:pso_p_update}\\
    \bv_n^{(i)} &= w^{(i)} \bv_n^{(i-1)} + c_1^{(i)} \br_1^{(i)} \left(\bbp - \bp_n^{(i)}\right) \notag\\
    &\quad + c_2^{(i)} \br_2^{(i)} \left(\bbbp^{(i)} - \bp_n^{(i)}\right),\label{eq:pso_v}\\\
    w^{(i)} &= \left(w_f - w_i\right) \cdot \left(i_\text{max}-i\right)/i_\text{max} + w_f,\label{eq:pso_w}\\
    c_1^{(i)} &= \left(c_{1f} - c_{1i}\right) i/i_\text{max} + c_{1i},\label{eq:pso_c1}\\
    c_2^{(i)} &= \left(c_{2f} - c_{2i}\right) i/i_\text{max} + c_{2i},\label{eq:pso_c2}
\end{align}
\end{subequations}

\subsubsection{Guaranteed Convergence PSO}
In the basic PSO formulation, the $\bar{n}^\text{th}$ particle that has the \textit{global best position} may become \textit{stagnant} when $\bp_{\bar{n}}^{(i)} = \bbp_n^{(i)} = \bbbp^{(i)}$, which leads to $\bv_{\bar{n}}^{(i)} = w^{(i)} \bv_n^{(i-1)} \to \mathbf{0}$. Consequently, this particle may stop searching for a better position, and the swarm tends to converge to $\bbbp^{(i)}$, which may not be a minimum. To avoid this phenomenon, we update $\bp_{\bar{n}}^{(i+1)}$ with the velocity given in \eqref{eq:gcpso}, where $n_s(i)$ and $n_f(i)$ are the number of successive successes and failures up to iteration $i$ (\cite{Engelbrecht2007}). A success is defined when $\ell\left(\bbbp^{(i+1)}\right) < \ell\left(\bbbp^{(i)}\right)$.
\begin{subequations}
    \begin{align}
        \bv_{\bar{n}}^{(i+1)} &= - \bp_{\bar{n}}^{(i)} + \bbbp^{(i)} + w^{(i)}\bv^{(i)} + \rho^{(i)} \left(1-2\br_2^{(i)}\right)\label{eq:gcpso_p}\\
        \rho^{(i+1)} &= \sysand{&2\rho^{(i)},& &n_s(i) > \epsilon_s,\\
                        &0.5\rho^{(i)},& &n_f(i) > \epsilon_f,\\
                        &\rho^{(i)},& &otherwise.}\label{eq:gcpso_rho}
    \end{align}\label{eq:gcpso}
\end{subequations}


\subsubsection{Termination conditions}
For the searching process to be terminated, the swarm \textit{cluster rate} $r_\text{clus}$ must be larger than a minimum value:
\begin{gather}
    r_\text{clus} = |\mathcal{C}|/N_\text{part} \ge r_\text{clus}^\text{min}.\label{eq:cluster_rate}
\end{gather}
The computing procedure for $r_\text{clus}$ can be found in the work of \cite{VanDenBergh2002}. We can avoid stopping the searching procedure prematurely by ensuring \eqref{eq:cluster_rate}. Once \eqref{eq:cluster_rate} holds, the searching process is terminated if one of the following conditions is satisfied:
\begin{subequations}
    \begin{gather}
        d\ell^{(i)} = \max_{j \in \set{i-N_\text{ter}, \dots, i}} \left|\bbbell^{(j)} - \bbbell^{(j-1)}\right|/\bbbell^{(i)} \le \epsilon_{d{\bbbell}},\\
        v_\text{max}^{(i)} = \max_{\substack{j \in \set{i-N_\text{ter}, \dots, i}\\n\in \set{0, N_\text{part}-1}}} \norm{\bv_n^{(j)}} \le \epsilon_{v},\\
        {\bbbell}(i) \le \epsilon_{{\bbbell}},\\
        i = i_\text{max}.
    \end{gather}\label{eq:termination}
\end{subequations}

Alg.~\ref{alg:pso_mpc} describes the overall computing procedure. Firstly, the search space, then at each iteration, we: i) re-initialize the swarm using Alg. \ref{alg:reinit_swarm} and Alg. \ref{alg:pso_constraints} to guarantee constraints; ii) apply the \textit{Global-best Guaranteed Convergence PSO (gbest-GCPSO)} method for convergence guarantees; iii) remove duplicated particles, and iv) check if terminating condition holds.

\begin{algorithm2e}[ht]\label{alg:pso_mpc}
  \SetAlgoLined
  
  \ForEach{$j\in \mathbb{N}_{0:N_p}$}{
  $\bp_{n,j}^{\text{ub},(i)}\gets  \max \left\{\bp^\text{max}, \bp_{n,j-1}^{(i),\text{ub}} + \Delta_p \right\}$\;
  $\bp_{n,j}^{\text{lb}, (i)}\gets \min \left\{\bp^{\text{min}}, \bp_{n,j-1}^{(i),\text{lb}} - \Delta_p \right\}$\;
  }
  $\bp_n^{(0)} \sim U\left(\bp^\text{lb}, \bp^\text{ub}\right)$; $\bv_n^{(0)} = \mathbf{0}$\;
  \For{$i \in \mathbb{N}_{1:i_\text{max}}$}{
    \lFor{$n\in \mathbb{N}_{0:N_\text{part}}$}{$\bp^{(i)} \gets $ Alg.~\ref{alg:reinit_swarm}}

    $\bbbp^{(i)}$, $\bbbell^{(i)}\gets$ \eqref{eq:pb_gb_position}; $\rho^{(i)}\gets $ \eqref{eq:gcpso_rho}\;
    
    \lFor {$n\in \mathbb{N}_{0:N_\text{part}}$}{$\bv_n^{(i)}\gets$ \eqref{eq:pso_v}}
    Remove duplicated particles\;
    \lIf{\eqref{eq:cluster_rate} \& any in \eqref{eq:termination} holds}
    {
        \textbf{return} $\bbbp^{(i)}$ \vspace{0.12cm}
    }
    
    }
  \caption{gbest-GCPSO-based NMPC}
\end{algorithm2e}

\subsection{Proposed PSO extensions for MPC and MHE}\label{sec:extended_pso}
In this subsection, we present the extensions tailored for this process dynamics after an in-depth analysis, as will be described later. We propose techniques for search-space initialization, swarm re-initialization, and duplicated particle removal. 

\subsubsection{Search-space initialization:} For the MPC optimization problem \eqref{eq:optimization_mpc} the search space can be limited based on the previous value of the input $u(k-1)$ (i.e., the initial guess), the input and its rate of change constraints \eqref{eq:mpc_u}-\eqref{eq:mpc_du}, $\bp^\text{min} := A_F^\text{min}+\mathbf{0}$, $\bp^\text{max} := A_F^\text{max}+ \mathbf{0}$, $\Delta_p := \Delta A_F^\text{max}+ \mathbf{0}$. Denote $\bp_{n,j}$ the $j^\text{th}$ element of the position vector of particle $n$, the search space, i.e., the particle position bounds can be computed as shown at the beginning of Alg.~\ref{alg:pso_mpc}. By narrowing the search space, we guarantee \eqref{eq:mpc_u}-\eqref{eq:mpc_du} and improve the optimum searching efficiency.


\subsubsection{Swarm re-initialization for constraints guarantee (Alg.~\ref{alg:reinit_swarm}):} 
\begin{figure}[H]
    \centering
    \includegraphics[width=\linewidth]{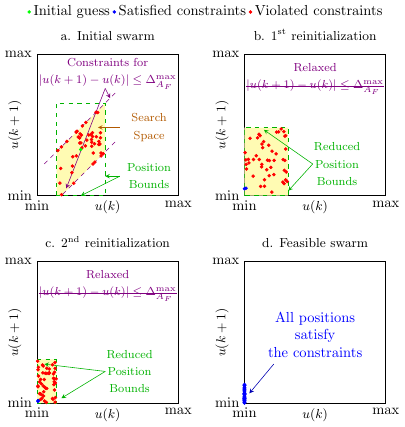}
    \caption{Particle swarm re-initialization to guarantee constraints at $t=4$ h in case (I), Sec.~\ref{sec:case_studies}.}
    \label{fig:adchem_pso_mpc}
\end{figure}

The state constraints \eqref{eq:mpc_x}-\eqref{eq:mpc_os} require more careful observations than those on control inputs, which can be guaranteed by applying saturation as detailed in Alg.~\ref{alg:pso_constraints}. Analyzing the process dynamics shows that guaranteeing on \eqref{eq:mpc_uaqD1}-\eqref{eq:mpc_os} can be achieved by reducing the control input magnitude. Therefore, reducing particle position can help reduce the constraint violation. To do this, we first try to reduce the velocity magnitude. Once the velocity is too small, and the constraints are not guaranteed, we relax the soft constraint \eqref{eq:mpc_du} and re-initialize the swarm with reduced bounds until: i) all particle positions satisfy \eqref{eq:mpc_uaqD1}-\eqref{eq:mpc_du}, or ii) $\bp^\text{lb} = \bp^\text{ub}=\bp^\text{min}$. This algorithm is illustrated in Fig.~\ref{fig:adchem_pso_mpc}, where the particles are shown in blue if constraints are guaranteed and in red if constraints are violated. The green dot denotes the initial guess of the optimization problem. We can see that the search space is narrowed after each time of re-initialization until all particles can guarantee constraints, $u(k) = A_F^\text{min}$ in Fig.~\ref{fig:adchem_pso_mpc}. 


\begin{algorithm2e}[ht]\label{alg:reinit_swarm}
  \SetAlgoLined
    \For {$n\in \mathbb{N}_{0:N_\text{part}}$}
    {
        $\bp_n^{(i)} \gets$ \eqref{eq:gcpso_p} and Alg.~\ref{alg:pso_constraints}\;
        
        \While{$\left\|\bv_n^{(i)}\right\| \neq 0$ and \textbf{not} \eqref{eq:mpc_uaqD1}-\eqref{eq:mpc_os}}{
            $\bv_n^{(i)} \gets \bv_n^{(i)}/2$\;
            $\bp_n^{(i)} \gets$ \eqref{eq:gcpso_p} and Alg.~\ref{alg:pso_constraints}\;
        }

        \While{ \eqref{eq:mpc_uaqD1}-\eqref{eq:mpc_os}}
        {
            \lIf{$\bp^\text{lb} \ge \bp^\text{min}$}{
                $\bp^\text{lb} \gets \bp^\text{lb} - \Delta_p$
            }
            \lElseIf{$\bp^\text{ub} \ge \bp^\text{min}$}{
                $\bp^\text{ub} \gets \bp^\text{ub} - \Delta_p$
            }
            \lElse{
                $\bp_n^{(i)} \gets \bp^\text{min}$
            }
        }
        $\bv_n^{(i-1)} \gets \bp_n^{(i)} - \bp_n^{(i-1)}$\;
    }    
  \caption{Swarm re-initialization}
\end{algorithm2e}

\subsubsection{Duplicated particles removal:} Duplicated particles are defined as those whose current position, best personal position, and velocity are the same. These particles will have the same behaviors in future iterations. Therefore, removing the duplicates will improve the computation cost and allow us to solve the optimization problem efficiently.

\begin{algorithm2e}[ht]\label{alg:pso_constraints}
  \SetAlgoLined
  $\bp_{n,-1}^{(i)} \gets u(k-1)$\;
  \For{$j\in \mathbb{N}_{0:N_p}$}{
  \lIf{$\bp_{n,j}^{(i)} < \bp^\text{lb}$}{
  $\bp_{n,j}^{(i)} \gets \bp^\text{lb}$
  }
  \lIf{$\bp_{n,j}^{(i)} > \bp^\text{ub}$}{
  $\bp_{n,j}^{(i)} \gets \bp^\text{ub}$}
  \lIf{$\bp_{n,j}^{(i)} - \bp_{n,j-1}^{(i)} > \Delta_{p}$}{$\bp_{n,j}^{(i)} \gets \bp_{n,j-1}^{(i)} + \Delta_{p}$}
  \lIf{$\bp_{n,j}^{(i)} - \bp_{n,j-1}^{(i)} < -\Delta_{p}$}{$\bp_{n,j}^{(i)} \gets \bp_{n,j-1}^{(i)} - \Delta_{p}$}
  $\bv_{n,j}^{(i-1)} \gets \bp_{n,j}^{(i)} - \bp_{n,j}^{(i-1)}$
  }
  \Return{$\bp_n^{(i)},\; \bv_{n,j}^{(i-1)}$}
  \caption{Guarantee position constraints.}
\end{algorithm2e}

\section{Case Studies}\label{sec:case_studies}
This section presents the simulation results of the developed control strategy. Firstly, we study the performance of the developed MPC on the full-state feedback nominal system. Secondly, we study the disturbed system with a $30\%$ decrease in $O_E$. For all simulations, we assume that the system is at a steady state at the beginning and contains only nitric acid and tributyl phosphate with nominal parameters, i.e., $\UaqF = 0$. Then, from $t=0$, $\UaqF$ is set to the nominal value. The set point is chosen as $\UaqD{9,\text{set}}=0.5 [U]^\text{aqD*}_9$ as shown in Fig.~\ref{fig:SSC}. Simulation parameters are given in Tab.~\ref{tab:simulation_params} and each simulation.

\subsection{Nominal case}
\begin{figure}[ht]
    \centering
    \includegraphics[width=\linewidth]{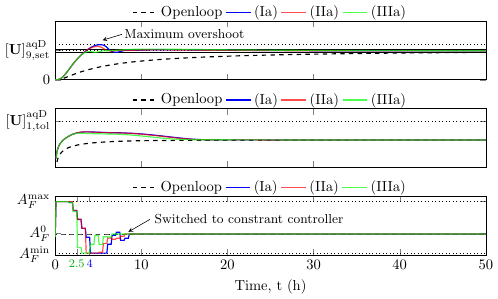}
    \caption{Simulation results for nominal PSO-based NMPC with $\bS=\bR=0.001/u_\text{set} \bI$.}
    \label{fig:adchem_1}
\end{figure}

\begin{figure}[ht]
    \centering
    \includegraphics[width=\linewidth]{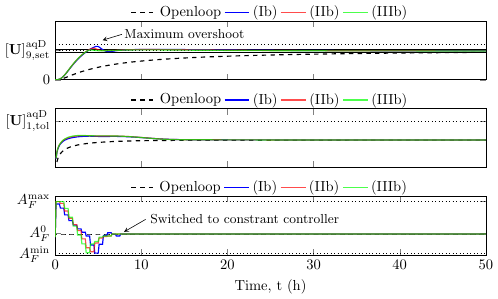}
    \caption{Simulation results for nominal PSO-based NMPC with $\bS=\bR=0.01/u_\text{set} \bI$.}
    \label{fig:adchem_2}
\end{figure}

Simulation results for nominal the PSO-based MPC are summarized in Tab.~\ref{tab:simulation_result}. As shown in Fig.~\ref{fig:adchem_1} and Fig.~\ref{fig:adchem_2}, all MPC formulations can effectively stabilize the system at the desired operating condition. It can be seen that there is a period (from 4 h to 6 h) that the Alg.~\ref{alg:reinit_swarm} is active to avoid constraint violations. In addition, it is shown that with a longer prediction horizon, MPC can better handle the constraints. For instance, in case (IIIa), MPC relaxes the soft constraint \eqref{eq:mpc_du} sooner compared to case (Ia) and (IIa) to avoid overshooting. As a result, this case has the lowest overshoot and settling time. Comparing Fig.~\ref{fig:adchem_1} and Fig.~\ref{fig:adchem_2} shows that a good trade-off in the cost function \eqref{eq:cost_function} also helps to avoid constraints violation and improve control performance. With larger values of $\bR, \bS$, \eqref{eq:optimization_mpc} return a solution that is closer to $u_\text{set}$. However, if $\bR, \bS$ are too large, \eqref{eq:optimization_mpc} will always return $u^*(k)=u_\text{set}$, which is in fact the open loop controller.



\begin{table}[ht]
\caption{Overall simulation parameters.}\label{tab:simulation_params}
\begin{center}
\begin{tabular}{@{}ll|ll@{}}
\toprule
\textbf{Parameter} & \textbf{Value} & \textbf{Parameter} & \textbf{Value} \\\midrule
$T_\text{meas}$           & 0.1 h                   & $N_\text{part}$       & 50                        \\ 
$T_\text{esti}$           & 0.5 h                   & $w_i$                 & 0.9                       \\
$T_\text{ctrl}$           & 0.5 h                   & $w_f$                 & 0.4                       \\
$\UaqD{9,\text{set}}$     & $0.5 [U]^\text{aqD*}_9$ & $c_{1i}=c_{2f}$       & 2.5                       \\
$\text{OS}^\text{max}$    & 20 \%                   & $c_{1f}=c_{2i}$       & 0.5                       \\
$\varepsilon_\text{ss}$   & 5 \%                    & $\Delta_p$ (MPC)      & $\Delta_{A_F}^\text{max}$ \\
$N_w$                     & 5                       & $\Delta_p$ (MHE)      & $\infty$                  \\
$N_e$                     & 2                       & $r_\text{clus}^\text{min}$   & 70 \%                     \\
$\varepsilon_{-N_e}$        &$0.1\% \UaqD{9,\text{set}}$& $N_\text{ter}$        & 5                         \\
$\varepsilon_\text{active}$& $1\% \UaqD{9,\text{set}}$ & $\epsilon_{\text{dl}}$ &1E-5\\
$N_{p2}$                   & 10                     & $\epsilon_{\text{v}}$  & 0.001\\
$\bQ = \bP$               & $\bI$                   & $\epsilon_{\text{l}}$  & 1E-4\\
                            &  & $i_\text{max}$         & 100                       \\ \bottomrule
\end{tabular}
\end{center}
\end{table}

\begin{table}[ht]
\caption{Results for PSO-based NMPC.}\label{tab:simulation_result}
\begin{center}
\begin{tabular}{@{}l|lll|lll@{}}
\toprule
Case          & (Ia) & (IIa) & (IIIa)& (Ib)& (IIb) & (IIIb)\\ \midrule
$\bS = \bR$ & \multicolumn{3}{c|}{$0.001/u_\text{set} \bI$} & \multicolumn{3}{c}{$0.01/u_\text{set} \bI$} \\
$N_p$         & 2 & 3 & 5            & 2 & 3 & 5            \\
Max OS        & 19.7\% & 13.5\% & 4.6\%        & 14.1 \% & 7.3 \% & 2.1 \%       \\
Settling time & 7.4 h& 5.7 h& 5 h          & 5.5 h & 4.8 h& 3.6 h        \\ \bottomrule
\end{tabular}
\end{center}
\end{table}

\subsection{Disturbed case}
\begin{figure}[ht]
    \centering
    \includegraphics[width=\linewidth]{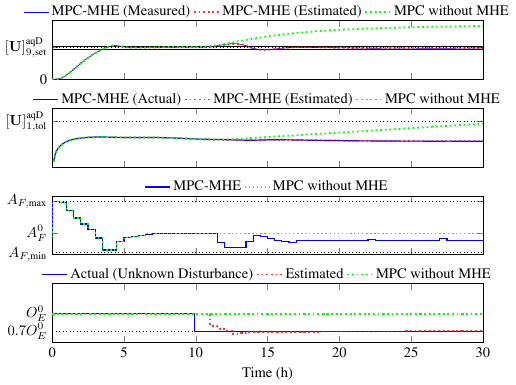}
    \caption{Simulation results for PSO-based adaptive NMPC under an unknown $30\%$ decrease of $O_E$.}
    \label{fig:adchem_sim_1}
\end{figure}

\begin{figure}[ht]
    \centering
    \includegraphics[width=\linewidth]{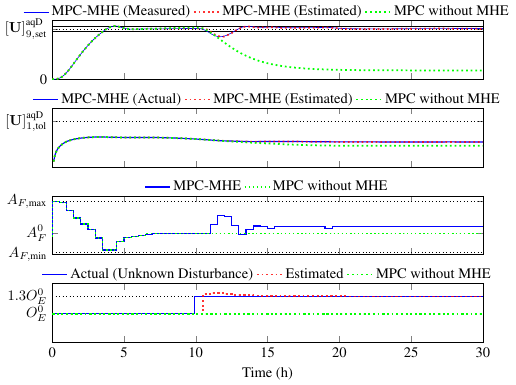}
    \caption{Simulation result for PSO-based adaptive NMPC under an unknown $30\%$ increase of $O_E$.}
    \label{fig:adchem_sim_2}
\end{figure}

In this subsection, we study the effects of disturbances on the system when a $\pm 30\%$ variation of $O_E$ (denoted as $p$ in \eqref{eq:state_space}) appears. We use the MPC with $N_p = 3$, $\bS=\bR=0.01/u_\text{set} \bI$ integrated with and MHE with parameters given in Tab.~\ref{tab:simulation_params}. The simulation results are shown in Fig.~\ref{fig:adchem_sim_1} and Fig.~\ref{fig:adchem_sim_2}, with $O_E^0$ as the nominal value of $O_E$ and the disturbances are assumed to appear at $t=10$ h. It can be seen that without MHE, MPC will continue to make predictions with its old parameter $O_E^0$. Consequently, the control input is kept constant, as in case (IIb). Thus, it fails to keep the output around the set point. In addition, even with MHE, it takes time for the effects of disturbances to appear in the measurements, i.e., $e_\text{esti}$ becomes sufficiently large. As a result, there is a short period just after $t=10$ h in which MPC does not predict the process evolution well since it keeps using the old value of $O_E$. This period results in an oscillation after $t=10$ h. However, thanks to the estimator, MHE, $O_E$ can be estimated and fed back to MPC. Finally, the controller can stabilize the system within the tolerance band \eqref{eq:steady_tolerance}.

\section{Conclusion}
This paper presents a nonlinear optimal adaptive control strategy for the uranium extraction-scrubbing operation in the PUREX process. The strategy was developed based on optimization-based control and estimation methods, i.e., nonlinear Model Predictive Control (NMPC) and nonlinear Moving Horizon Estimation (MHE). The methods were used for set-point tracking application to ensure the solvent saturation level of the process and parameter estimation since the solvent flow rate is an unknown disturbance. The NMPC and MHE optimization problems are solved using the Particle Swarm Optimization (PSO) algorithm, which allows the use of PAREX directly without requiring explicit mathematical equations. Multiple simulations were done using the PAREX code, the qualified simulation program that allows steady and transient calculations of the process. Simulation results show that the developed control strategy stabilizes the system at the desired operation condition, disregarding disturbances. Future developments, including studies on stability, robustness, uncertainties handling, and experimental implementation, will be studied in June 2024.

                                                   







\appendix
\end{document}